\documentclass[pdflatex,sn-nature]{sn-jnl}% 

\usepackage{graphicx}%
\usepackage{multirow}%
\usepackage{amsmath,amssymb,amsfonts}%
\usepackage{amsthm}%
\usepackage{mathrsfs}%
\usepackage[title]{appendix}%
\usepackage{xcolor}%
\usepackage{textcomp}%
\usepackage{manyfoot}%
\usepackage{booktabs}%
\usepackage{algorithm}%
\usepackage{algorithmicx}%
\usepackage{algpseudocode}%
\usepackage{listings}%
\usepackage{xspace}

\theoremstyle{thmstyleone}%
\theoremstyle{thmstyletwo}%

\theoremstyle{thmstylethree}%

\newcommand{\hi}{\textsc{H\,i}\xspace}

\begin{document}

\title[]{Neutral Atomic Hydrogen in a Star-forming Galaxy 7 Billion Years Ago}

\author*[1]{\fnm{Graham} \sur{Lawrie}}\email{grahamdavidlawrie@gmail.com}

\author[1,2]{\fnm{Roger} \sur{Deane}}\email{deane.roger@gmail.com}

\author[3,4]{\fnm{Tariq} \sur{Blecher}}\email{tariq.blecher@gmail.com}

\author[5,6]{\fnm{Danail} \sur{Obreschkow}}\email{danail.obreschkow@uwa.edu.au}

\author[7,8,3,4]{\fnm{Ian} \sur{Heywood}}\email{ian.heywood@skao.int}

\author[9]{\fnm{Shilpa} \sur{Ranchod}}\email{sranchod@mpifr-bonn.mpg.de}

\affil*[1]{\orgdiv{Wits Centre for Astrophysics}, \orgname{University of the Witwatersrand}, \orgaddress{\street{1 Jan Smuts Avenue}, \city{Johannesburg}, \postcode{2000}, \state{Gauteng}, \country{South Africa}}}

\affil[2]{\orgdiv{Department of Physics}, \orgname{University of Pretoria}, \orgaddress{\street{Lynnwood Rd, Hatfield}, \city{Pretoria}, \postcode{0002}, \state{Gauteng}, \country{South Africa}}}

\affil[3]{\orgdiv{Centre for Radio Astronomy Techniques and Technologies, Department of Physics and Electronics}, \orgname{Rhodes University}, \orgaddress{\street{Lucas Avenue}, \city{Makhanda}, \postcode{6140}, \state{Eastern Cape}, \country{South Africa}}}

\affil[4]{\orgname{South African Radio Astronomy Observatory}, \orgaddress{\street{Liesbeek House, River Park, Gloucester Road, Mowbray}, \city{Cape Town}, \postcode{7700}, \country{South Africa}}}

\affil[5]{\orgdiv{International Centre for Radio Astronomy Research (ICRAR), M468}, \orgname{University of Western Australia}, \orgaddress{\postcode{6009}, \state{WA}, \country{Australia}}}

\affil[6]{\orgname{ARC Centre of Excellence for All Sky Astrophysics in 3 Dimensions (ASTRO 3D)}}

\affil[7]{\orgname{SKA Observatory}, \orgaddress{Jodrell Bank, Lower Whitington, Macclesfield, SK11 9FT, UK}}

\affil[8]{\orgdiv{Astrophysics}, \orgname{University of Oxford}, \orgaddress{\street{Denys Wilkinson Building, Keble Road}, \city{Oxford}, \postcode{OX1 3RH}, \country{UK}}}

\affil[9]{\orgname{Max-Planck-Institute f\"ur Radioastronomie}, \orgaddress{\street{Auf dem H\"ugel 69}, \city{Bonn}, \postcode{53121}, \country{Germany}}}

\abstract{Neutral atomic hydrogen (\hi) constitutes a key phase of the cosmic baryon cycle, bridging the ionised circumgalactic medium and the star-forming molecular gas \cite{Sancisi_2008,Tumlinson_2017}. Yet, nearly 75 years after its discovery \cite{Ewen_1951}, direct views of \hi through its 21~cm emission line remain largely confined to the nearby Universe \cite{Zwaan_2005,Jones_2018}. Indirect measurements and statistical analyses indicate little evolution in the comoving \hi density over the past 10 billion years, in stark contrast to the order-of-magnitude decline in the cosmic star-formation rate density over the same epoch. Resolving this tension requires direct \hi measurements in individual, representative galaxies at earlier times. Here we report a detection of \hi 21~cm emission from the Dragon Arc, a gravitationally lensed main-sequence star-forming galaxy at $z=0.725$, observed 6.6 billion years in the past with the MeerKAT radio telescope. The inferred intrinsic \hi mass, $M_{\rm HI}=10^{9.66^{+0.16}_{-0.19}}\,\mathrm{M}_\odot$, and velocity width of $205^{+66}_{-48}\,\mathrm{km\,s^{-1}}$ are consistent with expectations from scaling relations for local star-forming galaxies\cite{Ponomareva_2021,Lelli_2019}. The resulting \hi depletion time of $1.2^{+1.0}_{-0.6}\,\mathrm{Gyr}$ is significantly shorter than the $\sim5-10$ Gyr, measured locally for comparable galaxies\cite{Saintonge_2017}. This indicates that the galaxy must rapidly replenish its atomic gas reservoir to remain on the star-forming main sequence. This detection demonstrates that strong gravitational lensing, combined with modern cm-wave facilities, can now reveal the \hi reservoirs of typical galaxies well beyond the local Universe, opening a new path toward statistical samples that will directly trace the evolution of the cosmic atomic gas supply.}

\maketitle

Abell~370 is a massive ($M_{500}\sim10^{15}\,\rm{M}_{\odot}$) galaxy cluster at redshift $z = 0.375$\cite{Struble_1999,Morandi_2007}. It is part of the Hubble Frontiers Fields (HFF) survey sample\cite{Lotz_2017}, a deep \textit{HST} and \textit{Spitzer} programme targeting massive galaxy clusters and leveraging gravitational lensing to study high-redshift galaxies as deep as 30-33 AB magnitudes.

\begin{figure*}
	\centering
    \includegraphics[width=0.7\columnwidth]{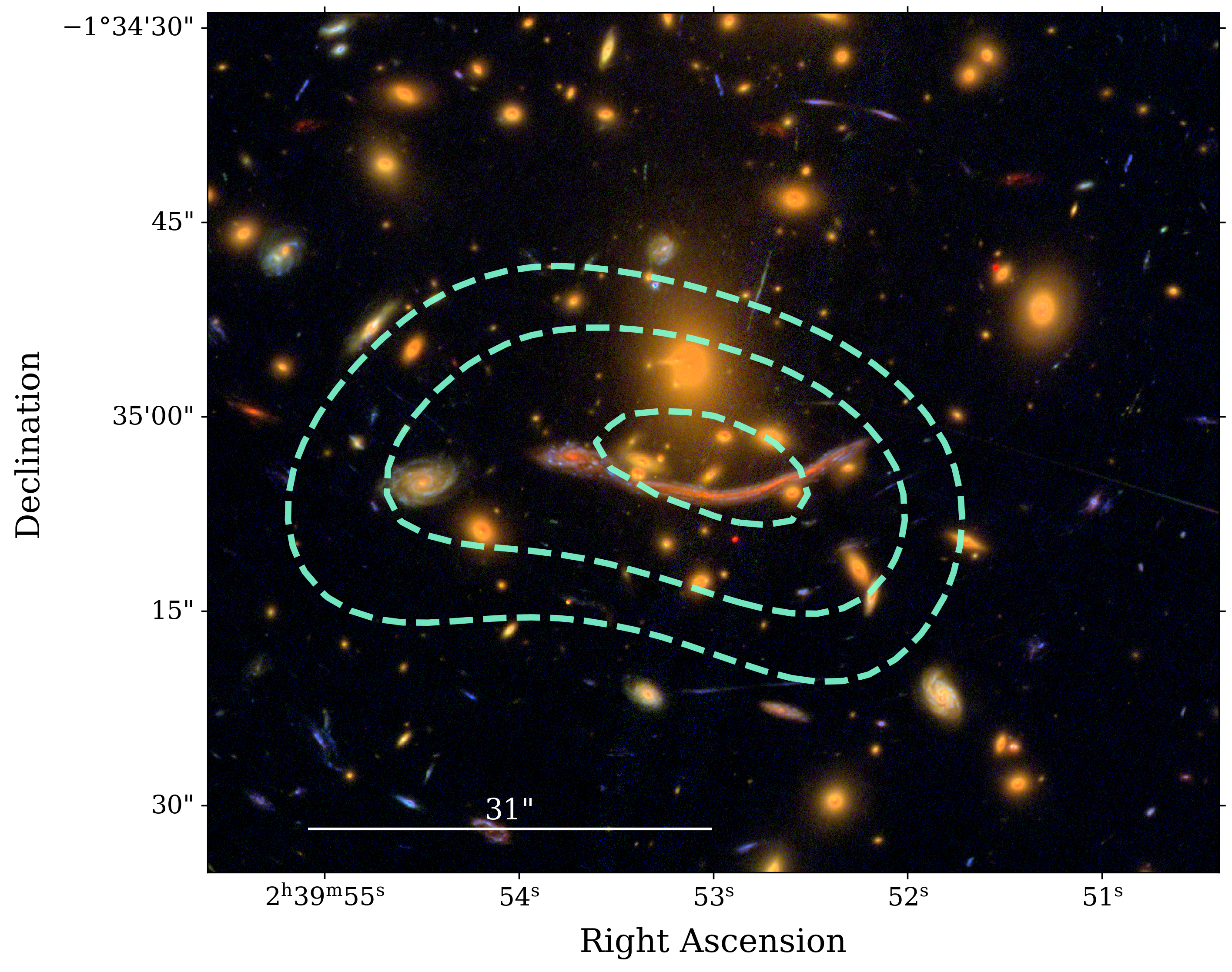}
    \caption{\textbf{Multiband \textit{HST} image of the Dragon Arc.} Overlaid on the multiband image (F814W, F606W, F435W) we show \hi contours at $134$, $143$, and $150\,\rm{Jy}\,\rm{Hz}\,\rm{beam}^{-1}$ ($4$, $4.25$, and $4.5\,\sigma$) from the total intensity \hi map generated from $4\times266\,\rm{kHz}$ channels. A 31" scale bar (bottom left) represents the semi-major axis of the smoothed total intensity map restoring beam.}
    \label{fig:arc_mom0}
\end{figure*}

The `Great Arc', also known more recently as known as the Dragon Arc\cite{Fudamoto_2025}, is a lensed star forming main sequence galaxy with a $\sim30$~arcsec optical extent approximately $25$~arcsec south of the foreground cluster centre of Abell~370. Fig~.\ref{fig:arc_mom0} shows a multi-colour \textit{HST} image of the Dragon Arc. It was first discovered in 1987\cite{Soucail_1987}, and since then followed up with deep \textit{HST}/ACS\cite{Richard_2010} and JWST\cite{Fudamoto_2025} programmes, modelling of which reveal a high optical/infrared (OIR) magnification ($\mu_{\rm{OIR}}=17\pm1$)\cite{Patricio_2018,Lagattuta_2017}. It has an intrinsic stellar mass $M_{\rm{*}}=10^{10.40\pm{0.01}}\,\rm{M}_\odot$\cite{Patricio_2018}, a mean star formation rate $\rm{SFR}=4.0^{+1.7}_{-0.9}\,\rm{M}_{\odot}\,\rm{yr}^{-1}$\cite{Patricio_2018}. This places the Dragon Arc on the star formation main sequence (SFMS)\cite{Speagle_2014,Patricio_2018}.

In Ref.~\cite{Blecher_2024} we investigated the prospects of direct detection of high-redshift lensed \hi emission behind the HFF clusters with \hi ray-tracing simulations using known, OIR-identified lensed galaxies. We predicted that multiple candidates are potentially detectable with MeerKAT within modest observation times. The most promising of these was the Dragon Arc, given its high predicted \hi magnification ($\mu_{\rm{HI}}=19\pm4$) and HI integrated flux ($S_{\rm{HI}}=119^{+70}_{-52}\,\rm{JyHz}$).

\section{Results}
We carried out a 10.3~hour on-source MeerKAT\cite{Jonas_2009,Jonas_2016,Camilo_2018} observation of the Dragon Arc using the Ultra-High Frequency (UHF) Band, spanning a frequency range of $544-1088\,\rm{MHz}$. Standard radio interferometric calibration is performed on a $\sim~40\,\rm{MHz}$ sub-band of two independent datasets using the semi-automated \textsc{oxkat}\cite{Heywood_2020} data processing pipeline, with particular care given to solving for direction-dependent gain corrections as well as accurate visibility domain subtraction of bright sources in the field that limit dynamical range and image fidelity. Continuum subtraction is performed in both the visibility and the image planes. We utilise a \textsc{Briggs robust=2.0} weighting scheme and $266\,\rm{kHz}$ ($\sim100\,\rm{km}\,\rm{s}^{-1}$) channel width when imaging spectral line \hi data cubes to maximise sensitivity. We use a beam sized aperture centred on the OIR Dragon Arc to extract the spectrum. Thereafter, we use nested sampling\cite{Skilling_2004} (\textsc{pymultinest}\cite{Buchner_2016}) for Bayesian parameter estimation and model selection (see Methods).
\begin{figure*}
    \centering
	\includegraphics[width=0.9\columnwidth]{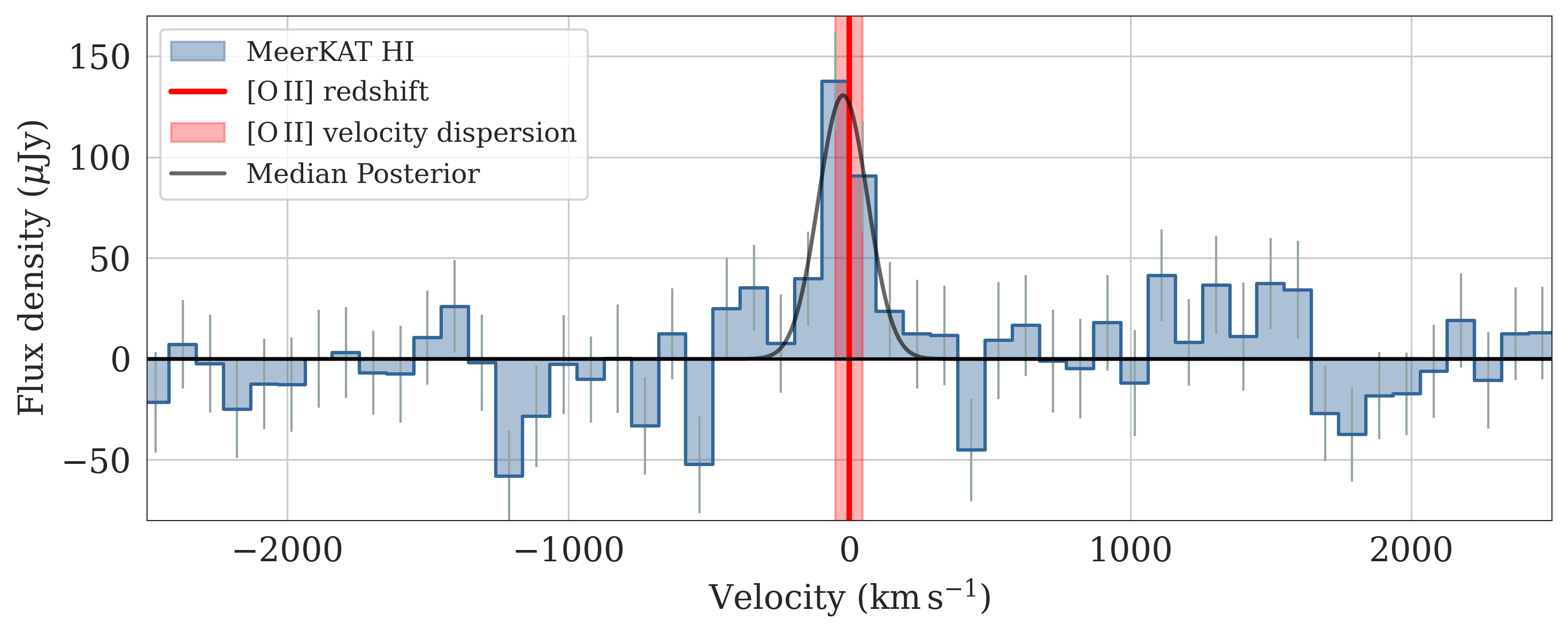}
    \caption{\textbf{MeerKAT neutral hydrogen (\textsc{HI}) emission spectrum towards the Dragon Arc.} The \hi centred at $z=0.7252$ is shown in blue. The thin red vertical line represents the optical spectroscopic redshift with the shaded red band corresponding to the velocity dispersion of the $[\rm{O}\,{II}]$ emission line\cite{Patricio_2018}. The median posterior of a single Gaussian model is shown in black. Grey vertical bars indicate the RMS of the flux distributions of 500 randomly placed identical apertures in each channel.}
    \label{fig:arc_spec2}
\end{figure*}

\begin{figure*}
    \centering
    \includegraphics[width=\textwidth]{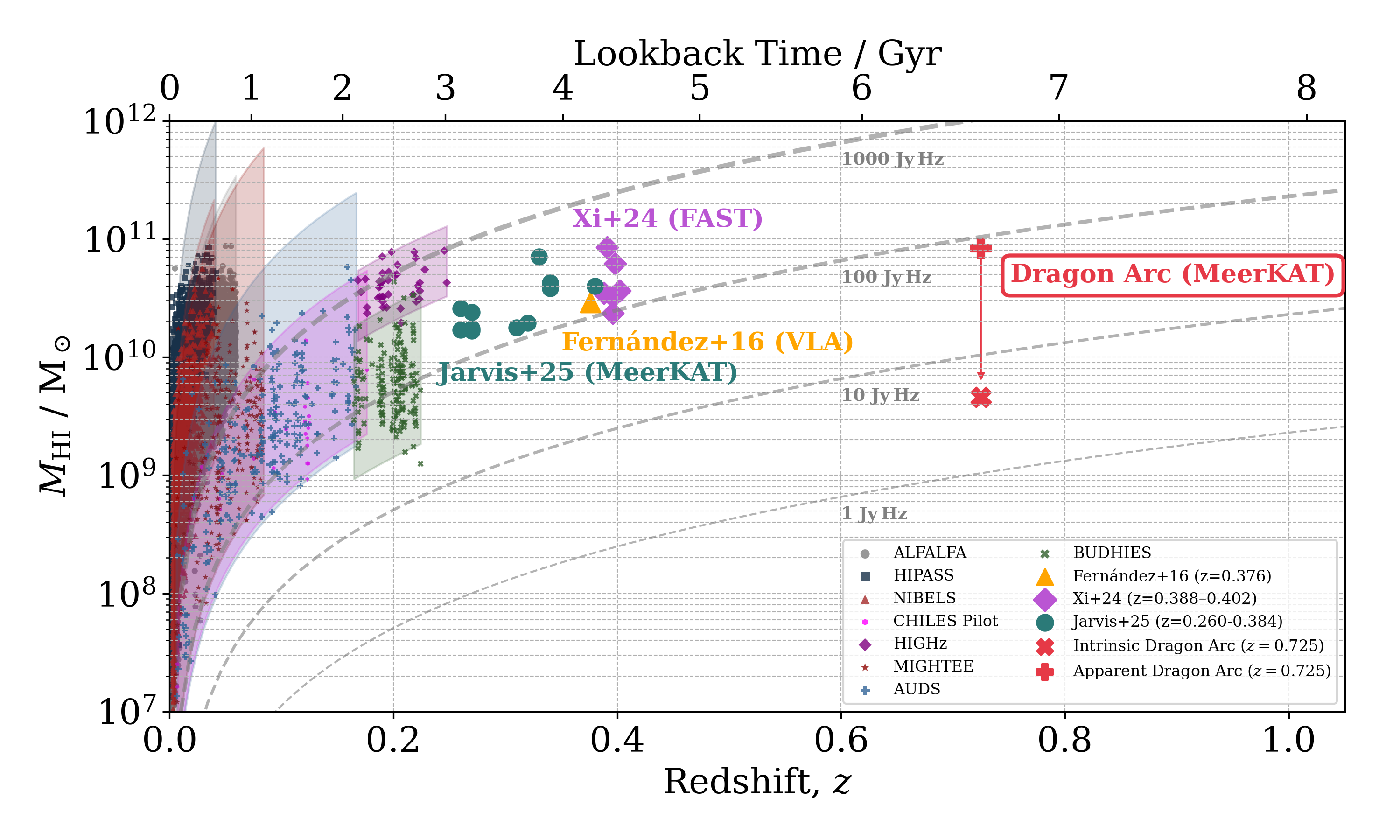}
    \caption{\textbf{\textsc{HI} mass as a function of redshift for direct detections.} The Dragon Arc detected in this work is shown in red. The VLA detection\cite[$z\sim0.376$,][]{Fernandez_2016} is shown by a yellow triangle, the FAST\cite[$z\sim0.4$,][]{Xi_2024} galaxies are shown with purple diamonds, and the MIGHTEE\cite[$z\sim0.25-0.4$,][]{Jarvis_2025} detections are shown with green circles. Shaded regions are defined by the 1 and 99 percentiles of the integrated fluxes of the samples for the respective surveys.}
    \label{fig:detection_distance}
\end{figure*}

The integrated \hi spectrum of the Dragon Arc, extracted with a beam sized aperture, is detected at an SNR of $5.6\,\sigma$ (Fig.~\ref{fig:arc_spec2}). Spectra in the two independent visibility datasets taken on different dates show peaks in the same channel, and the SNR increases when imaged together as the noise decreases by $\sim\sqrt{2}$, verifying the reliability of the detection. The centre of the single Gaussian model fit has a mean $\nu_{\rm{obs}}=823.34\pm0.05\,\rm{MHz}$, corresponding to $z_{\rm{HI}}=0.7252(1)$, while the $[\rm{O}\,{\rm{II}}]$-derived optical redshift is $z_{[\rm{O}\,{\rm{II}}]}=0.72505(1)$\cite{Patricio_2018}, shown with a red vertical line. These redshifts are consistent within velocity uncertainties of $\Delta {\rm{V}_{\rm{HI}}}=18\,\rm{km}\,\rm{s}^{-1}$ and $\Delta {\rm{V}_{[\rm{O}\,{\rm{II}}]}}=2\,\rm{km}\,\rm{s}^{-1}$, respectively. We generate a total intensity \hi map using a cube sub-band of $\sim400\,\rm{km}\,\rm{s}^{-1}$ rest-frame velocity width centred on the optical redshift $z_{\rm{[O\,{II}]}} = 0.72505(1)$\cite{Patricio_2018}, and smooth by a spatial Gaussian kernel with a semi-major axis matching the optical arc length of $\sim15\,\rm{arcsec}$. The Dragon Arc therefore corresponds to a near doubling of the highest redshift \hi emission direct detection to date, despite only requiring 10.3 hours of on-source integration time with MeerKAT (see Fig.~\ref{fig:detection_distance}). 

Attempts to model the \hi magnification map are limited by the large astrometric uncertainty ($\sim10$ arcsec) that results from our poor angular resolution ($\sim50$ arcsec). In contrast, the OIR-derived magnification from \textit{HST}/JWST is well constrained, tracing the stellar component has a modelled magnification of $\mu_{\rm{OIR}}=17\pm1$\cite{Patricio_2018,Lagattuta_2017}. When considering that the \hi is typically larger than the stellar extent\cite{Broeils_1997}, then a more accurate estimate may be to follow Ref.~\cite{Blecher_2024}, who use \hi mass-diameter and $M_{\rm{HI}}$-$M_{*}$ scaling relations to derive an expected magnification of $\mu_{\rm{HI}}=19\pm4$. What is clear from Ref.~\cite{Blecher_2024}'s statistical exploration of the magnification for a range of \hi profiles, inclinations, etc, is that the magnification is relatively well constrained between 10-20. Because of this, we adopt the mean of the OIR and \hi magnifications, $\mu_{\rm{HI}}=18^{+5}_{-3}$, with conservative uncertainties from Ref.~\cite{Blecher_2024}. We note that even a factor of 2-3 difference in this magnification would have no impact on scientific inferences of this paper. The measured integrated \hi flux is $S_{\rm{HI}} = 79.5^{+15.6}_{-14.3}\,\rm{JyHz}$ corresponding to an implied intrinsic \hi mass of ${M_{\rm{HI}}} = 10^{9.66^{+0.16}_{-0.19}}\, {\rm{M}_\odot}$. This is consistent with the predicted \hi mass, $M_{\rm{HI}} = 10^{9.73^{+0.14}_{-0.18}}\,{\rm{M}_\odot}$, assuming $M_{\rm{HI}}$-$M_{*}$ scaling relations and ray-tracing simulations\cite{Blecher_2024}. The velocity FWHM of the Dragon Arc is measured at $\rm{FWHM}= 205^{+66}_{-45}\,\rm{km}\,\rm{s}^{-1}$. These \hi, and magnification properties, as well as the aforementioned stellar mass and SFR properties, are summarised in Table~\ref {tab:arc_props}.

\begin{table*}
	\centering
	\caption{Dragon Arc measured and derived properties. }
	\label{tab:arc_props}
	\begin{tabular}{lc}
		\hline
		Property & Value\\
        \hline
		$ S_{\rm{HI}}$ & $79.53^{+15.58}_{-14.25}\,$JyHz\\
		$ \mu_{\rm{HI}}$ & $18^{+5}_{-3}$\\
		$\mu M_{\rm{HI}}$ & $10^{10.92^{+0.08}_{-0.09}}\,\rm{M}_\odot$\\
		$M_{\rm{HI}}$ & $10^{9.66^{+0.16}_{-0.19}}\,\rm{M}_\odot$\\
		$M^{\rm{pred}}_{\rm{H_2}}$ & $10^{9.57\pm0.20}\,\rm{M}_\odot$\\
		$M_*$ & $10^{10.40\pm0.01}\,\rm{M}_\odot$ \\
		$\rm{SFR}_{\rm{Balmer}}$ & $3.1\pm0.3\,\rm{M}_{\odot}\,\rm{yr}^{-1}$ \\
        $\rm{SFR}_{[\rm{O\,\rm{II}}]}$ & $3.1\pm0.6\,\rm{M}_{\odot}\,\rm{yr}^{-1}$ \\
        $\rm{SFR}_{\rm{SED}}$ & $5.68^{+0.20}_{-0.18}\,\rm{M}_{\odot}\,\rm{yr}^{-1}$ \\
		$z_{\rm{HI}}$ & 0.7252(1) \\
		$z_{\rm{[O\,\rm{II}]}}$ & 0.72505(1) \\
		$V_{\rm{offset}}$ & $22\,\rm{km}\,\rm{s}^{-1}$ \\
		$\rm{FWHM}$ & $205.4^{+66.1}_{-44.6}\,\rm{km}\,\rm{s}^{-1}$\\
        $t_{\rm{dyn}}$&$1.14^{+0.57}_{-0.41}\,\rm{Gyr}$\\
        $t_{\rm{dep,HI}}$&$1.16^{+0.98}_{-0.64}\,\rm{Gyr}$\\
		\hline
	\end{tabular}
\end{table*}

\section{Discussion}

\begin{figure*}
    \centering
    \includegraphics[width=\textwidth]{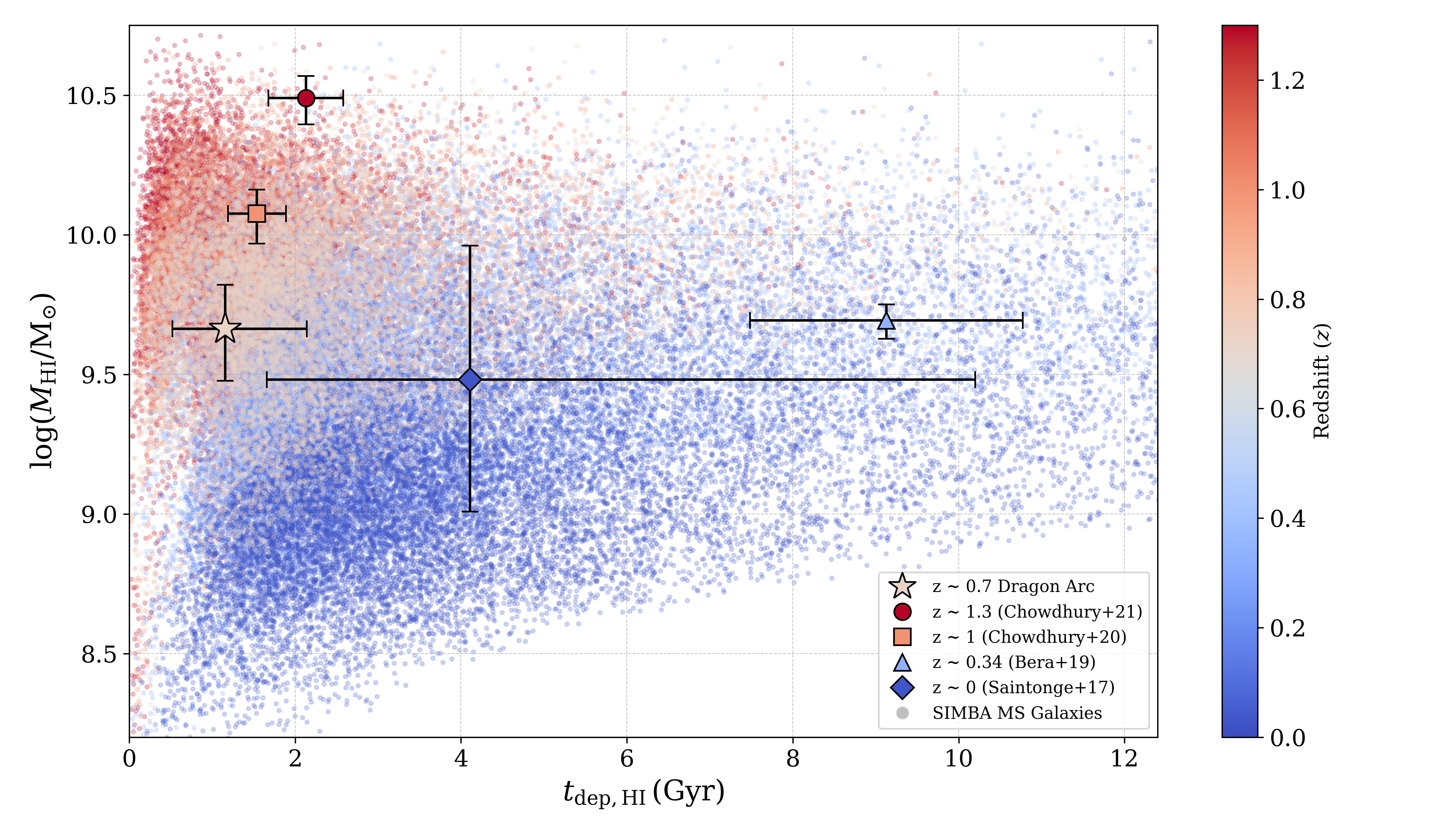}
    \caption{\textbf{Observational constraints on the \textsc{HI} gas depletion time in main-sequence galaxies across redshift space, compared with the \textsc{SIMBA} simulation.} Results from spectral-line stacking are shown with triangle, square, and circle markers\cite{Bera_2019,Chowdhury_2020,Chowdhury_2021}. The Dragon Arc result reported here is indicated with a star. We calculate a mean \textsc{HI} depletion time and \textsc{HI} mass for local SFMS galaxies from the xGASS sample\cite{Saintonge_2017,Catinella_2018}. SFMS galaxies from the flagship \textsc{SIMBA} simulation, $(100\,h^{-1}\,\rm{Mpc})^3$ box, are plotted with points. All points and markers are colourised by redshift.}
    \label{fig:t_dep}
\end{figure*}

A key question in galaxy evolution is how galaxies on the SFMS sustain star formation, which is commonly investigated through measuring the \hi depletion time, $t_{\rm{dep,HI}}=M_{\rm{HI}}/\rm{SFR}$. Given the derived $M_{\rm{HI}}$ and mean estimated $\rm{SFR}$, we calculate a \hi depletion timescale of $\rm{t}_{\rm{dep},\rm{HI}}= 1.16^{+0.98}_{-0.64}\,\rm{Gyr}$ for the Dragon Arc. The average \hi depletion timescales from stacking results are consistent with that of the Dragon Arc, at $\langle\rm{t}_{\rm{dep,\rm{HI}}}\rangle=1.54\pm0.35\,\rm{Gyr}$ and $\langle\rm{t}_{\rm{dep,\rm{HI}}}\rangle=2.13\pm0.45\,\rm{Gyr}$, reported for $\sim7600$ and $\sim2000$ stacked galaxies at $z\sim1.0$ and $z\sim1.3$, respectively\cite{Chowdhury_2020,Chowdhury_2021}. Similar to Ref.~\cite{Chowdhury_2020,Chowdhury_2021} the \hi depletion time for the Dragon Arc is significantly lower than the local \hi depletion timescale range, $5-10\,\rm{Gyr}$, depending on stellar mass distribution\cite{Saintonge_2017,Bera_2019}. In Fig.~\ref{fig:t_dep} we contextualise the \textsc{HI} depletion time of the Dragon Arc by comparing it to \hi stacking results\cite{Bera_2019,Chowdhury_2020,Chowdhury_2021}, local SFMS xGASS galaxies\cite{Saintonge_2017,Catinella_2018}, and SFMS galaxies from the cosmological hydrodynamical simulation, \textsc{SIMBA}\cite{Dave_2019}. We select galaxies as SFMS galaxies if they lie within $0.3\,\rm{dex}$ of the SFMS\cite{Speagle_2014}. This reinforces the assertion that, without significant \hi replenishment, the Dragon Arc is an example of a SFMS galaxy that will not have a sufficiently large cold gas reservoir to sustain star-formation to $z\sim0$, consistent with what higher redshift ($z \gtrsim 0.7$) \hi stacking results suggest for typical SFMS galaxies.

\begin{figure*}
    \centering
    \includegraphics[width=0.9\columnwidth]{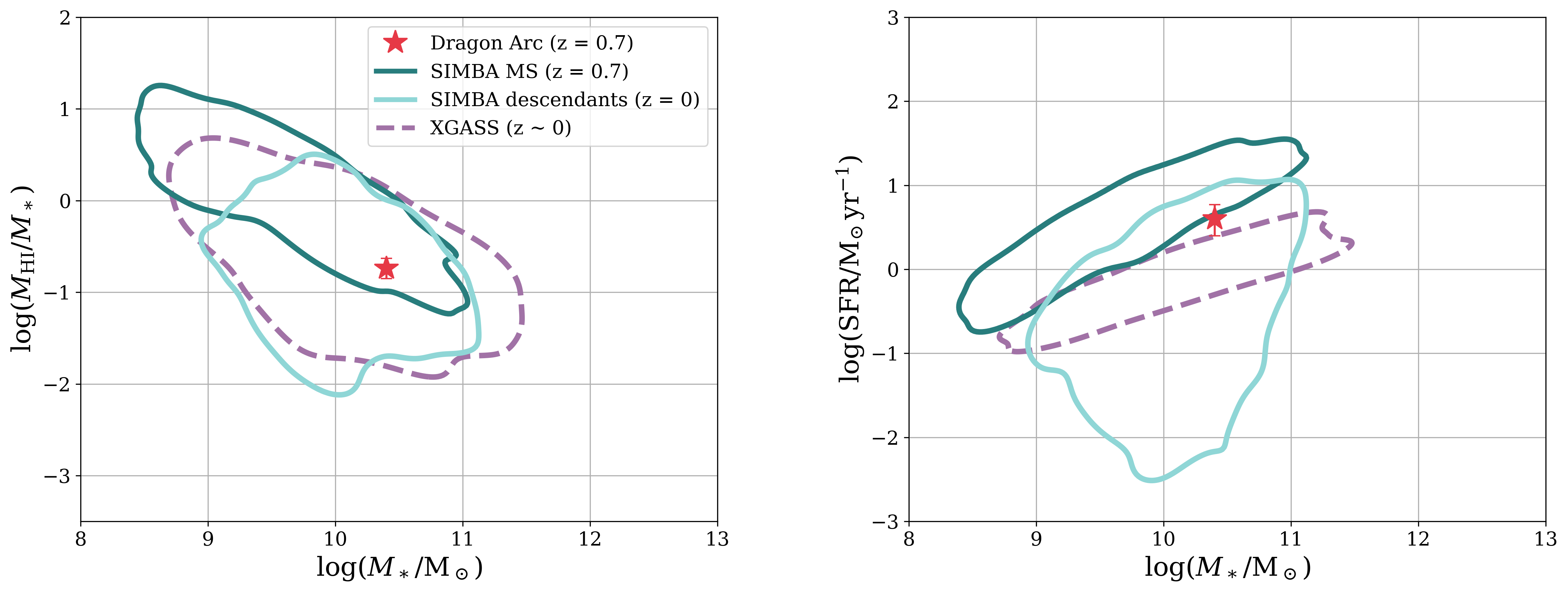}
    \caption{\textbf{Scaling relation comparison between local observations and the \textsc{SIMBA} cosmological hydrodynamical simulations.} \hi-to-stellar mass fraction (left) and SFR (right) as a function of stellar mass. Cyan and teal contours represent \textsc{SIMBA} main sequence galaxies at $z=0$ and $z=0.7$ respectively, where the former are the individually tracked descendants of the latter. The dashed purple contours represent galaxies in the xGASS ($z\sim0$) sample\cite{Catinella_2018}. The red star indicates the estimated intrinsic properties of the Dragon Arc. Contours enclose $95\,\%$ of the data.}
    \label{fig:simba_xgass}
\end{figure*}

To further explore its potential evolutionary paths, we investigate the Dragon Arc properties relative to established scaling relations and the SFMS between $z=0.7$ and $z=0$. The Dragon Arc is compared to SFMS galaxies in the xGASS $z\sim0$ sample\cite{Catinella_2018}, and simulated SFMS galaxies and their descendants from $z=0.7$ to $z=0$ using \textsc{SIMBA}\cite{Dave_2019}. This choice is motivated by the shape and characteristic \hi mass ($M^{\star}_{\rm{HI}}$) of the \hi mass function (HIMF) being accurately reproduced with $M_{\rm{HI}}\gtrsim10^9\,\rm{M}_\odot$ \textsc{SIMBA} galaxies at $z\sim0$\cite{Jones_2018,Dave_2020}. Fig.~\ref{fig:simba_xgass} shows the \hi-to-stellar mass fraction (left) and SFR (right) as a function of stellar mass. Descendants of SFMS galaxies identified at $z=0.7$ are individually tracked to $z=0$ using the \texttt{progen} module in \textsc{CAESAR}\footnote{https://caesar.readthedocs.io/en/latest/}. The Dragon Arc, indicated with a red star, is consistent with the $z=0.7$ SFMS \textsc{SIMBA} $95\,\%$ contours in terms of $M_{\rm{HI}}$, $M_{*}$, and $\rm{SFR}$. The \textsc{SIMBA} galaxies' descendents contours closely match contours of the observed galaxies in the xGASS sample at $z\sim0$ in terms of $M_*$ and $M_{\rm{HI}}$ (Fig.~\ref{fig:simba_xgass} left panel). However, when considering the SFMS (Fig.~\ref{fig:simba_xgass} right panel), \textsc{SIMBA} descendents are seen to bifurcate from the xGASS contours, with $58\,\%$ departing from the SFMS and evolving towards the quiescent `red cloud'\cite{Strateva_2001}, while the rest of the galaxies remain on the SFMS. These two distinct evolutionary pathways illustrate the importance of directly measuring the \hi content of individual galaxies out to the lookback times corresponding to $\gtrsim6$ dynamical timescales. These comparisons demonstrate the Dragon Arc is consistent with simulated SFMS galaxies, which then evolve to have \hi-to-stellar mass fraction consistent with low-redshift measurements. We investigate the effect of feedback mechanisms on the aforementioned bifurcation by comparing results from different \textsc{SIMBA} variants (`no X-ray', `no jet', `no AGN'). We find that radiative AGN winds and jets are primarily responsible for quenching low and high-stellar mass galaxies, respectively (see Methods). This result highlights the importance of commensal, high-resolution radio continuum data to map these radio jets alongside deep \hi observations from MeerKAT and SKA-MID, arrays designed to detect \hi emission out to $z\sim1$.  

The Dragon Arc \hi detection validates predictions that gravitational lensing can be leveraged to enable high-redshift direct detections of \hi emission within modest integration times. This result demonstrates that large-scale targeted programmes of massive galaxy clusters at appropriately selected redshifts with MeerKAT UHF and SKA-MID Band 1, together with statistical wide field lensed \hi searches\cite{Deane_2015,Button_2025}, will deliver statistically significant samples of lensed \hi detections\cite{Blecher_2019,Ranchod_2022,Blecher_2024}. These samples will enable important comparisons with indirect methods such as stacking and intensity mapping while complementing deep \hi surveys such as LADUMA\cite{Blyth_2016}. Together, these complementary \hi programmes will elucidate our view of this key missing component in intermediate to high-redshift samples to fully understand the baryonic cycle and its evolution over cosmic time.

\section{Methods}
\label{sec:methods}

\subsection{Cosmological Parameters}
\label{sec:cosmo}

Throughout this work we use a flat $\Lambda$-cold dark matter ($\Lambda$CDM) cosmology with $\rm{H}_0=67.66\,\rm{km}\,\rm{s}^{-1}\,\rm{Mpc}^{-1}$ and $\Omega_{\rm{m}} = 0.31$\cite{Planck_2020}. Using this, 1 arcsecond corresponds to $7.36\,\rm{kpc}$ at $z=0.7$. 

\subsection{Observation and data processing}
\label{sub:data}

We use MeerKAT observations taken in the UHF band ($544-1088\,\rm{MHz}$), project ID: SCI-20210212-TB-01. These were carried out on 2021-09-13/14 and 2021-10-04/05 with the correlator in the wideband fine (32k) channelisation mode, with a native spectral resolution of $16.6\,\rm{kHz}$. The correlator dump time was set to 8~sec. The observation was split into two separate full tracks of $\sim6$ hours each, resulting in a total of $\sim 10.3$ hours on-source integration time of Abell~370, centred on $\rm{RA}=02\rm{h}39\rm{m}53\rm{s}$ and $\rm{Dec}=-01\rm{d}35\rm{m}00\rm{s}$. Both tracks had 61 of the total 64 MeerKAT antennas participate. For further technical details, see \cite{Jonas_2009,Jonas_2016,Camilo_2018}.

These MeerKAT raw visibilities are reduced using \textsc{Oxkat}\footnote{https://github.com/IanHeywood/oxkat}\cite{Heywood_2020}, a semi-automatic data reduction pipeline that implements standard data reduction routines including flagging; bandpass, absolute flux density, and complex gain reference calibration (1GC); self-calibration (2GC); and direction-dependent calibration (3GC). \textsc{Oxkat} utilises \textsc{Casa}\footnote{https://casa.nrao.edu}\cite{McMullin_2007}, \textsc{WSClean}\cite{Offringa_2014}, \textsc{CubiCal}\footnote{https://github.com/ratt-ru/CubiCal}\cite{Kenyon_2018}, \textsc{DDFacet}\footnote{https://github.com/saopicc/DDFacet}\cite{Tasse_2023_facet}, and \textsc{killMS}\footnote{https://github.com/saopicc/killMS}\cite{Tasse_2023_killms} among other software, and is fully described in Ref.~\cite{Heywood_2020,Heywood_2022,Heywood_2024}.

The following procedure applies to both MeerKAT UHF Abell 370 datasets, with minor variations in the manual calibration steps. We extract 2200 channels at the native $16.6\,\rm{kHz}$ frequency resolution, corresponding to a bandwidth of $\sim37\,\rm{MHz}$ centred on $\sim823.4\,\rm{MHz}$. Known radio frequency interference (RFI) regions are excised, and lower-level RFI is identified through several algorithms. Autoflagging is performed using \textsc{CASA}'s \texttt{flagdata} task in \texttt{rflag} and \texttt{tfcrop} modes to remove RFI. Minor manual flagging is then applied to eliminate any remaining radio interference. To correct for atmospheric phase shifts and other effects, we apply standard reference calibration procedures using \texttt{CASA} tasks such as \texttt{gaincal} and \texttt{applycal}. A final iteration of flagging is performed using the autoflagger \textsc{Tricolour}\footnote{https://github.com/ratt-ru/tricolour}. Using \textsc{WSClean} together with \textsc{CubiCal}, we perform self-calibration to enhance image fidelity and dynamic range by iteratively refining the mask output by \textsc{WSClean}.

In both measurement sets, direction-dependent calibration was essential due to problematic bright off-axis sources, beyond and within the primary beam FWHM. We explore several self-calibration strategies, finding that the following performs best. We sequentially peel the two brightest off-axis sources, where peeling is the process of solving for complex gains on the visibilities phase rotated towards the relevant problematic source. We do so using \textsc{CubiCal} and thereafter subtract the derived model from the visibilities. After peeling, we re-image centred on the original pointing centre to inspect the results and find a significant reduction in contaminating ripples from the now subtracted bright contaminant sources.

The visibilities are then imaged with \textsc{DDFacet} after redetermining the cleaning mask on the higher dynamic range image. We select eight sources to initiate the tessel region determination for which direction-dependent complex gains are solved for using the \textsc{killMS} package. Imaging these corrected visibilities shows a significant improvement in the image quality and dynamic range. The final continuum image RMS is $\sigma \sim 20~\mu$Jy\,beam$^{-1}$ for both of the independent datasets, consistent with the expected thermal noise given the $\sim40\,\rm{MHz}$ bandwidth used. All imaging in the direction-independent and dependent stages of self-calibration uses a Briggs {\sc robust} = -0.5 parameter setting, which has corresponding PSF FWHM of $12.08''$.

Continuum emission was subtracted as follows in a 3-step process consistent with well established MeerKAT best practices. First, in the visibility plane, we subtract the final continuum sky model. We then fit and subtract a first-order polynomial to the visibilities, using the full 8-second time resolution to identify further continuum emission. Spectral imaging is performed using \textsc{WSClean}. We image 100 channels at an averaged frequency resolution of $266\,\rm{kHz}$ and spatial dimensions of $5120\times 5120$ pixels, with a pixel scale of $2$~arcsec per pixel. We perform various tests of \textsc{WSClean} parameters including but not limited to: \texttt{-auto-threshold}/\texttt{-threshold}, \texttt{-auto-mask}/\texttt{-fits-mask}, and \texttt{-circular-beam}. We find that any chosen \textsc{WSClean} strategy, using reasonable thresholds ($0.3-2.5\,\sigma$) and masking techniques, lead to at most a $\sim10\,\%$ difference in flux. This is significantly lower than the \hi flux uncertainty (Sec.~\ref{sub:fitting}). To enhance our sensitivity to fainter \hi emission, we adopt a \textsc{Briggs robust=2} weighting scheme. We perform same suite of tests on a \textsc{Briggs robust=0.5} scheme, limiting our sensitivity in favour of improved angular resolution, and find the same consistency with varying \textsc{WSClean} parameters, albeit at a lower SNR ($\sim4$), as one would expect when down-weighting shorter baselines sensitive to $\gtrsim30$ arcsec scales. Following spectral cube generation, we use \textsc{Casa}'s \texttt{imcontsub} to fit a fourth-order polynomial to the continuum emission in the image plane across the $37\,\rm{MHz}$. Attempts with lower order polynomials do not result in any significant changes in the vicinity of the \hi emission line. This image-plane continuum fit serves as our model for subtracting further residual continuum emission, which is mostly present at larger off-axis angles, well away from our science target. The final data cube has a PSF FWHM of $54.1''$.

\subsection{HI spectrum modelling}
\label{sub:fitting}

\begin{figure*}
	\includegraphics[width=\columnwidth]{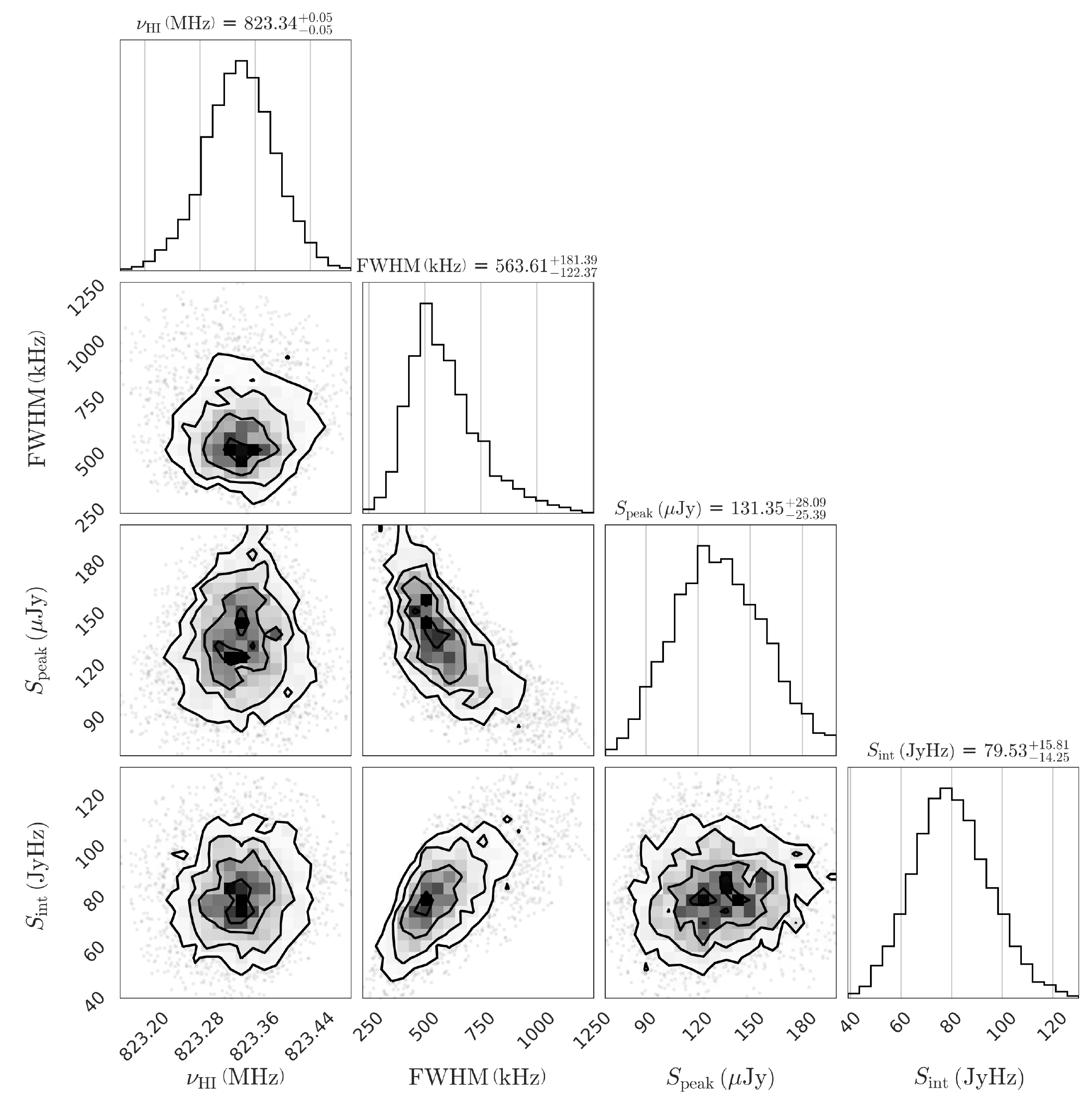}
    \caption{\textbf{\textsc{Robust=2} posteriors probability distribution functions.} For the single Gaussian model, we show distributions for central frequency ($\nu_{\rm{HI}}$), velocity FWHM, and peak flux density ($S_{\rm{peak}}$). Additionally, we show the distribution for the derived integrated flux density, $S_{\rm{int}}$.}
    \label{fig:corner2}
\end{figure*}

We use nested sampling\cite{Skilling_2004} in the \textsc{pymultinest}\footnote{https://github.com/JohannesBuchner/PyMultiNest}\cite{Buchner_2016} package for Bayesian model selection and parameter estimation of the \hi emission line properties. In Fig.~\ref{fig:corner2} we show the posterior probability distribution functions of a single Gaussian model of the emission line, including the parameters: peak amplitude ($S_{\rm{peak}}$), $\rm{FWHM}$, and central frequency ($\nu_{\rm{HI}}$), while the integrated flux ($S_{\rm{int}}$) is a derived parameter.

To estimate flux uncertainties in the integrated \hi spectrum shown in Fig.~\ref{fig:arc_spec2}, we place 500 beam sized apertures randomly within a bounding box near, but not overlapping with, the Dragon Arc region and calculate the flux within each aperture by summing the pixel values and normalising by the beam factor, thereby returning a flux density distribution for each channel. The standard deviation of each distribution is used as a per-channel uncertainty (shown as grey vertical uncertainties in Fig.~\ref{fig:arc_spec2}).

Regardless of the approach taken to determine the SNR of the detection, we find an $\rm{SNR}\gtrsim5$. The ratio of amplitude to spectrum RMS in Fig.~\ref{fig:arc_spec2} yields an SNR of $5.6\,\sigma$, while the integrated SNR derived from the integrated flux density posteriors uncertainties in Fig.~\ref{fig:corner2} is $5.3\,\sigma$. The median posteriors used to derive all \hi properties are shown in Fig.~\ref{fig:corner2}.

\subsection{HI properties}
\label{sub:hi_prop}

\begin{figure*}
	\includegraphics[width=\columnwidth]{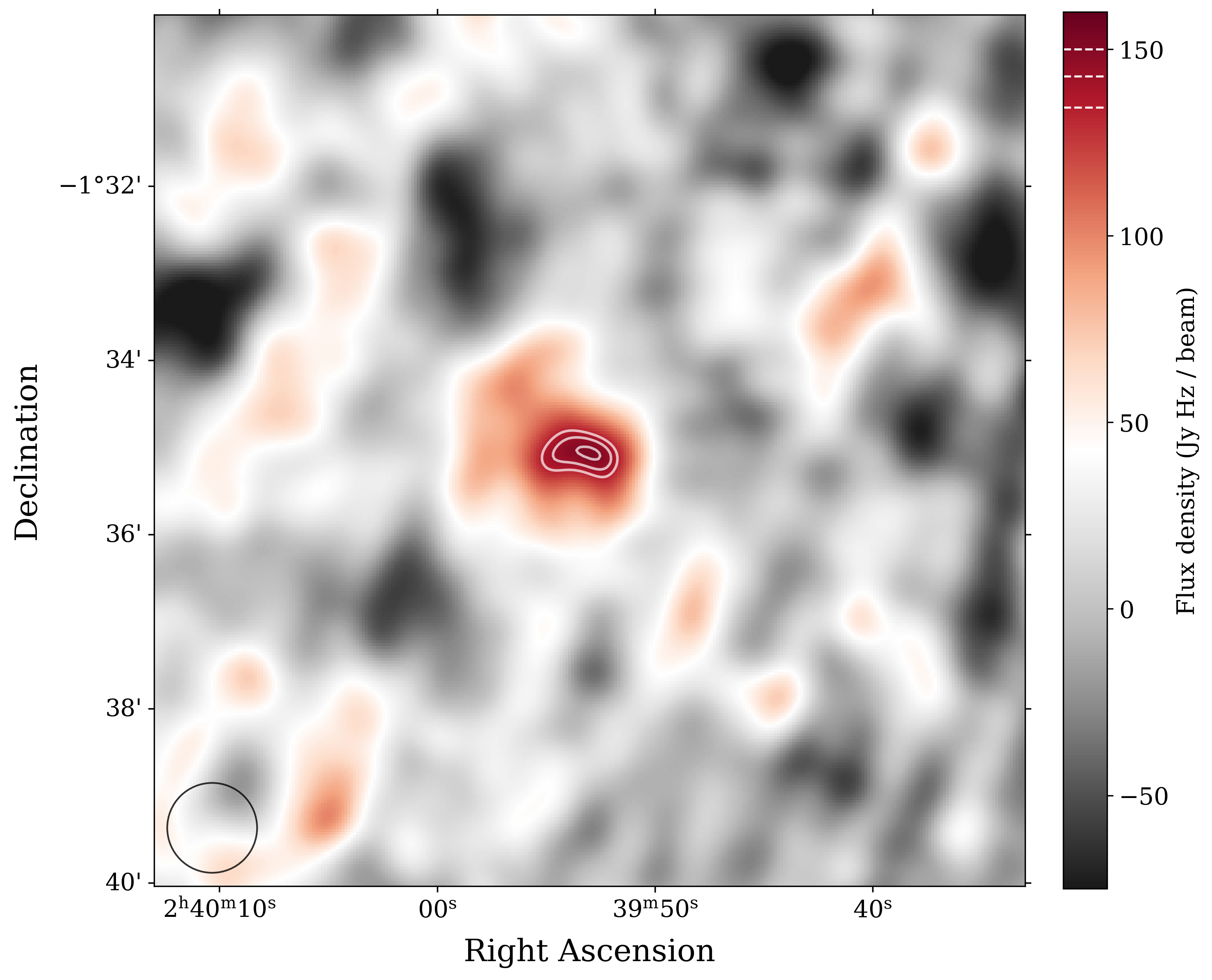}
    \caption{\textbf{\textsc{HI} total intensity map of the Dragon Arc.} The map is generated from 4 \hi channels and smoothed by the OIR extent sized Gaussian with semi-axes of 15 arcseconds. In the bottom left we show the smoothed PSF FWHM, $62''$ arcsec.}
    \label{fig:mom0_zoomedout}
\end{figure*}

As outlined, \hi properties are derived using the parameter posteriors of a single Gaussian model applied to the spectral line data cube (see Fig.~\ref{fig:corner2}). To derive the \hi mass we adopt the method in Ref.~\cite{Meyer_2017},
\begin{equation}
    \left( \frac{\mu_{\rm{HI}} M_{\rm{HI}}}{\rm{M}_\odot} \right) = 49.7 \left(\frac{D_{\rm{L}}}{\rm{Mpc}} \right)^2 \left( \frac{S}{\rm{JyHz}} \right),
\end{equation}
where $M_{\rm{HI}}$ is the \hi mass in $\rm{M}_{\odot}$, $S$ is the apparent (magnified) integrated flux of the \hi emission in $\rm{JyHz}$, and $D_{\rm{L}}$ is the luminosity distance to the source in $\rm{Mpc}$. With a minor modification that accounts for the magnification of the \hi flux, represented by $\mu_{\rm{HI}}$. Using the spectrum extracted from a beam sized aperture centred on the OIR Dragon Arc, together with $z_{\rm{HI}}$, we find an apparent \hi mass of $\mu M_{\rm{HI}}=10^{10.92^{+0.08}_{-0.09}}\,{\rm{M}_\odot}$, and a magnification corrected intrinsic mass of $M_{\rm{HI}}=10^{9.66^{+0.16}_{-0.19}}\,{\rm{M}_\odot}$. For context, we show a total intensity (moment 0) \hi map integrated over $4\times266\,\rm{kHz}$ channels ( $\Delta\rm{V}\sim400\,\rm{km}\,\rm{s}^{-1}$) with the same contours as Fig.~\ref{fig:arc_mom0} but displaying a larger scale in Fig.~\ref{fig:mom0_zoomedout}.

\subsection{Dynamical timescale derivation}
\label{sub:dynamical_mass}

We estimate the \hi rotational velocity, correcting for inclination, as
\begin{equation}
    v_{\rm{rot}} = \frac{W}{2 \sin(i)},
\end{equation}
where $v_{\rm{rot}}$ is the rotational velocity in $\rm{km}\,\rm{s}^{-1}$, $W=\rm{FWHM}=205.36^{+66.15}_{-44.61}\,\rm{km}\,\rm{s}^{-1}$ is the width of the line profile between the approaching and receding side of the galaxy, and $i=75\pm5\,\rm{degrees}$\cite{Blecher_2024} is the inclination angle of the galaxy. We find $v_{\rm{rot}}=106.30^{+38.17}_{-24.69}\,\rm{km}\,\rm{s}^{-1}$ for the Dragon Arc. Using the \hi diameter-mass relation\cite{Wang_2016},
\begin{equation}
    \log\left(\frac{D_{\rm{HI}}}{\rm{kpc}}\right) = (0.506 \pm 0.003) \log\left(\frac{M_{\rm{HI}}}{\rm{M}_\odot}\right) - (3.293 \pm 0.009),
\end{equation}
where $D_{\rm{HI}}$ is the diameter of the \hi disk defined where the surface density is $1\,\rm{M}_\odot\,\rm{pc}^{-1}$, we find a corresponding $D_{\rm{HI}}=39.49^{+5.85}_{-5.10}\,\rm{kpc}$ with uncertainties from the $0.06\,\rm{dex}$ scatter. We then assume the \hi disk radius, defined as $R_{\rm{HI}}=D_{\rm{HI}}/2$, is a valid proxy to determine the dynamical time, 
\begin{equation}
    t_{\rm{dyn}} \approx \frac{2\pi R_{\rm{HI}}}{v_{\rm{rot}}},
\end{equation}
to find $t_{\rm{dyn}}=1.14^{+0.57}_{-0.41}\,\rm{Gyr}$.

\subsection{Stellar and molecular gas properties}
\label{sub:stel_prop}

The intrinsic stellar mass, $M_{\rm{*}}=10^{10.40\pm0.01}\,{\rm{M}_\odot}$\cite{Patricio_2018}, associated with the Dragon Arc lensed system was calculated by fitting both spectra and photometry using FSPS (Flexible Stellar Population Synthesis)\cite{Conroy_2010} and \textsc{Prospector}\footnote{https://prospect.readthedocs.io/en/stable/}\cite{Johnson_2021}. The $\rm{SFR}$ for the system is taken to be $\rm{SFR}=3.97^{+1.7}_{-0.9}\,\rm{M}_{\odot}\,\rm{yr}^{-1}$, which was calculated as the mean of the three $\rm{SFR}$s derived by Ref~.\cite{Patricio_2018}, $\rm{SFR}_{\rm{[\rm{O}\,{\rm{II}}]}}=3.1\pm0.6\,\rm{M}_{\odot}\,\rm{yr}^{-1}$, $\rm{SFR}_{\rm{Balmer}}=3.1\pm0.3\,\rm{M}_{\odot}\,\rm{yr}^{-1}$, and $\rm{SFR}_{\rm{SED}}=5.68^{+0.20}_{-0.18}$ with uncertainties from the range of $\rm{SFR}$ values. The Balmer line estimate uses the Ref~.\cite{Kennicutt_1998a} calibration and the [$\rm{O}\,{\rm{II}}$] line uses the Ref.~\cite{Kewley_2004} calibration. The authors note that magnification is corrected by a flux scaling of $1/\mu$ on a spaxel-by-spaxel basis, but using an averaged magnification factor leads to a negligible difference. They make use of the magnification model from Ref.~\cite{Lagattuta_2017} and obtain a mean magnification for the Dragon Arc system of $\mu_{\rm{OIR}}=17\pm1$. We estimate the molecular gas content of the Dragon Arc by assuming the $\rm{SFR}-M_{\rm{H_2}}$ relation\cite{Sargent_2014} for non-starburst galaxies to find $M_{\rm{H_2}} =10^{9.57\pm0.20}\,\rm{M}_\odot$, where the uncertainties correspond to the scatter of the relation and helium has been corrected for.

\subsection{Scaling relations with varying AGN feedback models in \textsc{SIMBA}}
\label{sub:ms_comp}

Here we classify a galaxy as a SFMS galaxy when $\Delta \rm{SFR}<0.3 \,\rm{dex}$, where $\Delta \rm{SFR} = |\rm{SFR} - \rm{SFR}(\rm{SFMS})|$ and $\rm{SFR}(\rm{SFMS)}$ is the expected $\rm{SFR}$ for a given stellar mass\cite{Speagle_2014}.

We turn to simulations to explore the evolutionary fate of SFMS galaxies and their modelled \hi gas at $z=0.7$. The \textsc{SIMBA} simulation\cite{Dave_2019} is a cosmological hydrodynamical simulation with unique black hole growth and multi-modal feedback. We utilise the following different feedback variations available in the $(50\,h^{-1}\,\rm{Mpc})^3$ box:
\begin{itemize}
  \item `no X-ray' turns off X-ray feedback while leaving jet and radiative AGN winds on
  \item `no jet' turns off both X-ray and jet feedback
  \item `no AGN' turns off X-ray, jet, and radiative AGN winds leaving only stellar and supernova feedback
\end{itemize}

\begin{figure*}
	\includegraphics[width=\columnwidth]{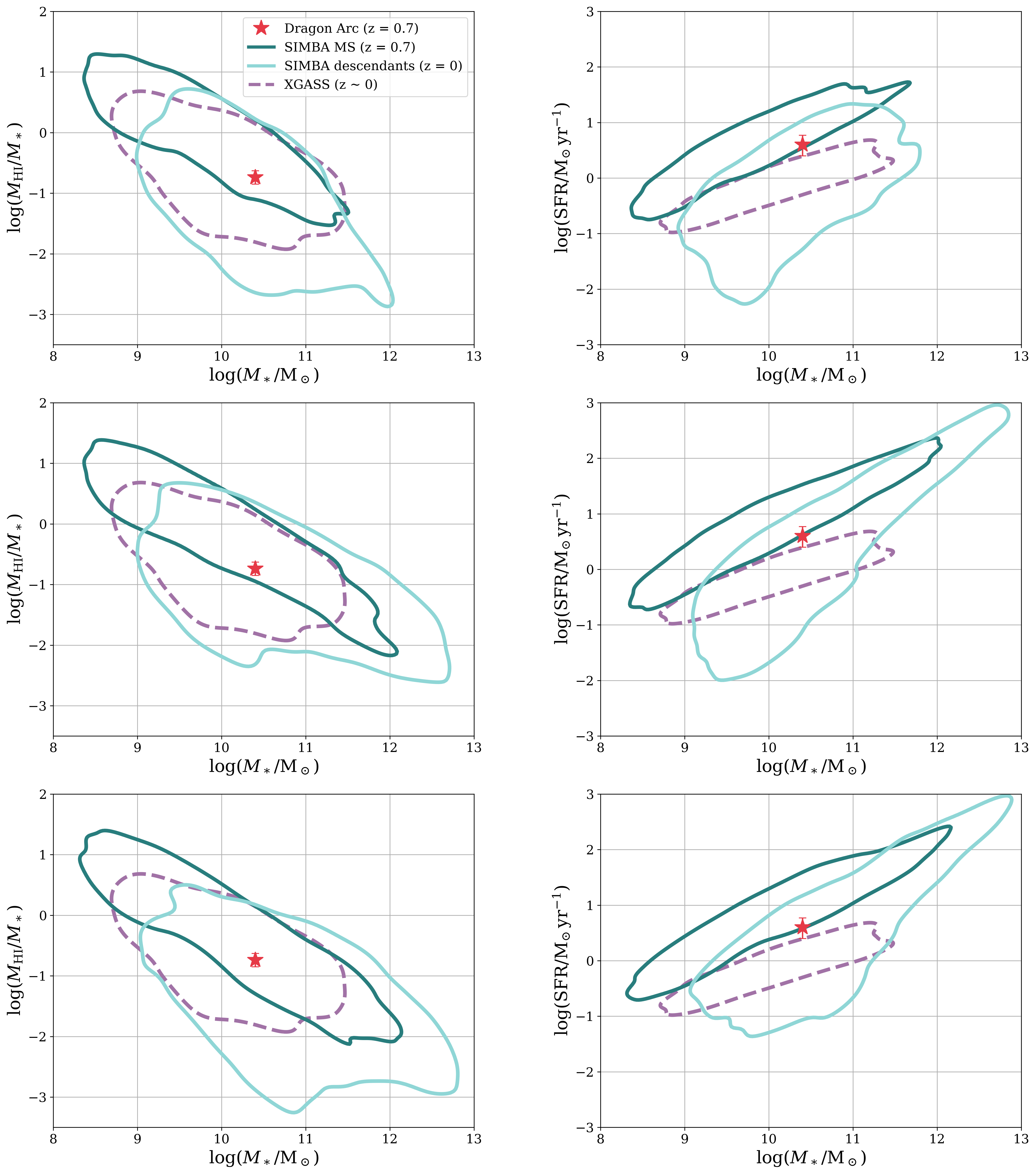}
    \caption{\textbf{Scaling relation comparison between local observations and \textsc{SIMBA} for differing feedback simulations.} \hi-to-stellar mass fraction (left) and SFR (right) as a function of stellar mass. From top to bottom, the feedback variants are as follows: no X-ray feedback, no jets or X-ray feedback, and no AGN feedback whatsoever, respectively. Cyan and teal lines represent \textsc{SIMBA} main sequence galaxies at $z=0$ and $z=0.7$ respectively, where the former are the individually tracked descendants of the latter. The dashed purple line represents galaxies in the xGASS ($z\sim0$) sample\cite{Catinella_2018}. The red star indicates the estimated intrinsic properties of the Dragon Arc. Contours enclose $95\,\%$ of the data.}
    \label{fig:simba_xgass_noagn}
\end{figure*}

To investigate the bifurcation between the `blue cloud' and `red cloud' seen in Fig.~\ref{fig:simba_xgass}, we repeat the descendant tracking but for \textsc{SIMBA} feedback variant snapshots (Fig.~\ref{fig:simba_xgass_noagn}). The `no X-ray' run (top row) results in marginal changes when compared to the full feedback run seen in Fig.~\ref{fig:simba_xgass}. In the `no jet' run, the number of SFMS galaxies that shift towards quiescence is reduced significantly, from 58 to $21\,\%$. In the `no AGN' run (bottom row), this is further reduced to only $9\,\%$ of \textsc{SIMBA} galaxies move off of the SFMS between $z=0.7$ and $z=0$. Fig.~\ref{fig:simba_xgass_noagn} highlights the known role of radiative AGN winds in quenching low mass ($M_{*}\leq10^{10}\,\rm{M}_\odot$) galaxies, while AGN jets are primarily responsible for quenching high mass ($M_{*}\geq10^{11.5}\,\rm{M}_\odot$) SFMS galaxies\cite{Dave_2019}.

It is worth noting that with all feedback mechanisms toggled on (Fig.~\ref{fig:simba_xgass}), galaxies that shift off of the SFMS towards quiescence are replaced by galaxies that join the SFMS. On the flagship \textsc{SIMBA} run, $(100\,h^{-1}\,\rm{Mpc})^3$, we find that while $58\,\%$ of the 13145 SFMS galaxies identified at $z=0.7$ shift towards quiescence, other galaxies move onto the SFMS, making for a total of 20309 SFMS galaxies at $z=0$. 

\subsection{Data Availability}

The measurement sets used in this work are available at the MeerKAT archive\footnote{https://archive.sarao.ac.za} under project ID: SCI-20210212-TB-01.

\bibliography{formatted_output}

\end{document}